# Pressure-driven dome-shaped superconductivity and electronic structural evolution in tungsten ditelluride

Xing-Chen Pan[1,2,*], Xuliang Chen[2,3,4,*], Huimei Liu[1,2,*], Yanqing Feng[1,2], Zhongxia Wei[1,2], Yonghui Zhou[2,4], Zhenhua Chi[2,3,4], Li Pi[2,3], Fei Yen[4], Fengqi Song[1,2], Xiangang Wan[1,2], Zhaorong Yang[2,3,4], Baigeng Wang[1,2], Guanghou Wang[1,2] & Yuheng Zhang[2,3]

Tungsten ditelluride has attracted intense research interest due to the recent discovery of its large unsaturated magnetoresistance up to 60 T. Motivated by the presence of a small, sensitive Fermi surface of $5d$ electronic orbitals, we boost the electronic properties by applying a high pressure, and introduce superconductivity successfully. Superconductivity sharply appears at a pressure of 2.5 GPa, rapidly reaching a maximum critical temperature ($T_c$) of 7 K at around 16.8 GPa, followed by a monotonic decrease in $T_c$ with increasing pressure, thereby exhibiting the typical dome-shaped superconducting phase. From theoretical calculations, we interpret the low-pressure region of the superconducting dome to an enrichment of the density of states at the Fermi level and attribute the high-pressure decrease in $T_c$ to possible structural instability. Thus, tungsten ditelluride may provide a new platform for our understanding of superconductivity phenomena in transition metal dichalcogenides.

[1] National Laboratory of Solid State Microstructures, College of Physics, Nanjing University, Nanjing 210093, China. [2] Collaborative Innovation Center of Advanced Microstructures, Nanjing University, Nanjing 210093, China. [3] High Magnetic Field Laboratory, Chinese Academy of Sciences, Hefei, Anhui 230031, China. [4] Key Laboratory of Materials Physics, Institute of Solid State Physics, Chinese Academy of Sciences, Hefei, Anhui 230031, China. * These authors contributed equally to this work. Correspondence and requests for materials should be addressed to F.S. (email: songfengqi@nju.edu.cn) or to X.W. (email: xgwan@nju.edu.cn) or to Z.Y. (email: zryang@issp.ac.cn).

As a new class of two-dimensional (2D) materials, transition metal dichalcogenides (TMDs) $MX_2$, where M is a transition metal (Mo, W, Re and so on) and X is a chalcogen (S, Se and Te), have attracted tremendous attention due to their rich physics and promising potential applications[1–12]. Sizable band gaps that can change from indirect to direct in single layers have been found in many TMDs, including $WS_2$, $WSe_2$, $MoS_2$ and $MoSe_2$. This property allows for the development of nanosized electrical transistors, and of electroluminescent and photodetector devices[4]. Field-effect transistors using thin films of TMDs as channel materials are found to exhibit an extremely high on–off current ratio[6]. Charge-density waves and superconductivity have also been observed in TMDs[13,14]. Of particular interest is the dome-shaped superconducting phase observed in a gate-tuned $MoS_2$ device, and this effect is also commonly seen in many unconventional superconductors[15].

Most recently, an extremely large positive magnetoresistance (MR) was discovered at low temperatures in non-magnetic tungsten ditelluride ($WTe_2$) TMD[1]. In contrast to other materials, the MR of $WTe_2$ remains unsaturated even at extremely high applied magnetic fields of 60 T[1]. It has been observed that at low temperatures the hole and electron pockets are approximately the same size[1,3], and that disruption in the balance between the two results in strong suppression of the MR[2]. A perfect balance between the electron and hole populations may therefore be the primary source of these novel and unwavering MR effects[1–3]. As a semimetal[1,3], the density of states at the Fermi level ($N(E_F)$) is rather low[16], and no superconductivity has ever been detected down to 0.3 K (ref. 2).

High pressure has been shown to be a clean and powerful means of generating novel physical states[17], having been particularly effective not only in tuning the $T_c$ of the superconductivity in elements[18] and compounds[19,20] but also in inducing superconductivity with ferromagnetic or antiferromagnetic orders as their ground states at ambient pressure. Moreover, in the case of $WTe_2$, the Te-5p and W-5d orbitals are spatially extended, thus making it very sensitive to variations caused by external pressure and strain. This property could shed some light on the high-pressure induction of superconducting transport in $WTe_2$. Here we observe the pressure-induced superconductivity, which exhibits a critical temperature ($T_c$) of 7 K at the pressure of 16.8 GPa. A dome-shaped $T_c$–P phase diagram is demonstrated. It is interpreted by the theoretical calculations.

## Results

**Pressure-induced superconductivity.** As shown in Fig. 1a, $WTe_2$ is a layered TMD material and the layer stacking results in a unit cell with four formula units and orthorhombic symmetry (its space group is $Pnm2_1$)[21]. The Te–Te bonds between the Te–W–Te sandwich layers are weak; therefore, nanoflakes with thicknesses down to several nanometres can be exfoliated using a scotch tape-based mechanical method. Moreover, the W atoms form zigzag chains along the a axis resulting in a one-dimensional substructure within a 2D material[1]. We grew single crystals of $WTe_2$ using a vapour transport method, in which the structural parameters were obtained by X-ray diffraction. As seen in the Supplementary Tables 1 and 2, $WTe_2$ exhibits similar atomic structures to those reported previously[21]. At ambient pressure, the resistivity decreases smoothly with decreasing temperature as shown in Fig. 1b. No phase transition was observed down to 2 K. Figure 1c shows the magnetic field-dependent transport measured for various temperatures at ambient pressure, which confirms the large MR reported recently[1,2].

Figure 2a,b shows the evolution of the resistance as a function of temperature in a single crystal of $WTe_2$ at various pressures.

The pressure was increased from 2.5 to 16.1 GPa in run no. 1 as shown in Fig. 2a, and from 9.3 to 68.5 GPa for another crystal in run no. 2 as shown in Fig. 2b. In run no. 1, at a pressure of 2.5 GPa, the resistance decreased monotonically with decreasing temperature, exhibiting typical metallic behaviour (inset of Fig. 2a). A superconducting transition was observed at $T_c = 3.1$ K. Here we define $T_c$ to be the onset temperature at which the drop in resistance occurs. At pressures up to 16.1 GPa, $T_c$ increased with increasing applied pressure. In the pressure curves for 2.5, 4.9 and 8.45 GPa, zero resistance was not seen because the superconducting transition was not complete at the lowest temperatures achievable using our equipment. If one check Fig. 2a carefully, it can be seen that the broadening widths at higher pressures are around 2 K. If such broadening keeps in the low pressure <8.45 GPa, the zero-resistance temperature will be extended below 1.8 K, which is the low limit of our equipment. At 11 GPa, however, zero resistance was observed for some temperatures. In the case of run no. 2, with pressures starting at 9.3 GPa, $T_c$ first increased slightly to a maximum of 7 K at 16.8 GPa, where it began to decrease monotonically with increasing pressure as shown in Fig. 2b. A suppression of two to three orders of magnitude of the MR emerged once superconductivity appeared, as evidenced by the MR curves at 10 K shown in Fig. 2c. To demonstrate that the zero resistance represented superconductivity, we also performed a.c. susceptibility measurements as shown in Fig. 2d, where the diamagnetic signal was observed at 8.3 and 12.5 GPa. This is in good agreement with the resistance measurements. The onset of MR suppression and superconductivity under pressure were thus demonstrated.

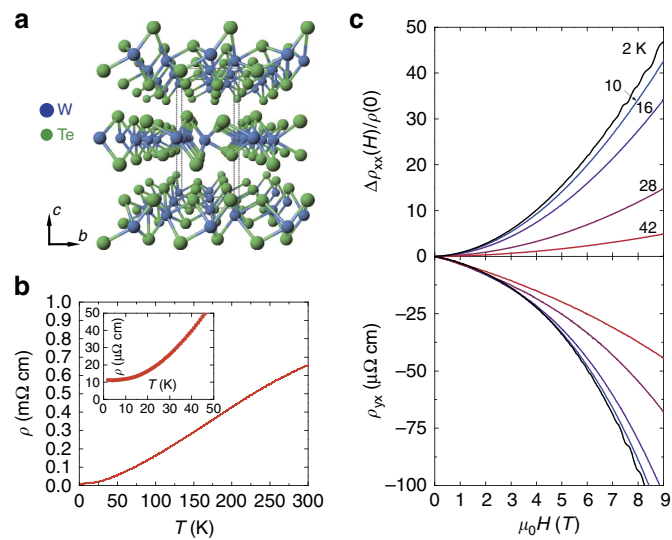

**Figure 1 | $WTe_2$ transport measurements at ambient pressure.** (**a**) The atomic structure of the $WTe_2$ crystal. Blue and green circles represent W and Te, respectively. (**b**) Temperature dependence of electrical resistivity at ambient pressure. The inset shows detail of data below 50 K with no hint of any superconductivity. (**c**) The magnetoresistance (upper plot) and Hall resistivity (down plot) at different temperatures at ambient pressure. Different colours represent different temperatures as marked.

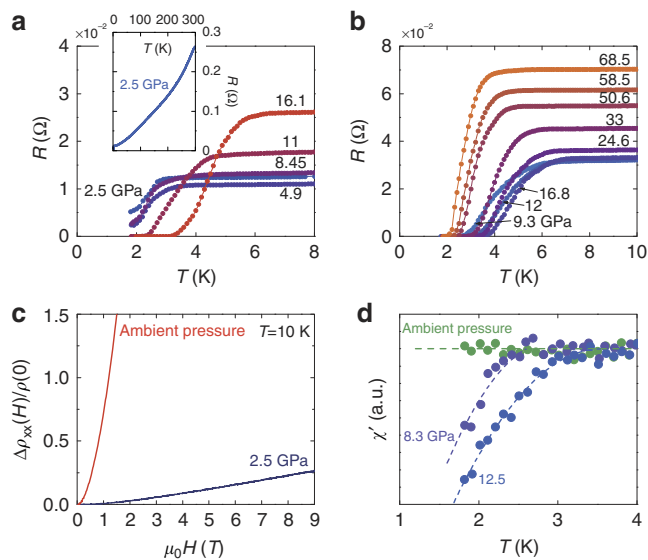

**Figure 2 | Experimental evidence of pressure-induced superconductivity.** (**a**) The temperature-dependent resistance under different pressures up to 16.1 GPa in run no. 1. The inset shows the temperature-dependent resistance from 1.8 to 300 K at 2.5 GPa. The onset of superconductivity can be seen from the drop in resistance. (**b**) Temperature dependence of resistance under various pressures from 9.3 to 68.5 GPa in run no. 2. (**c**) Magnetoresistance comparison at 10 K between ambient pressure and 2.5 GPa. Magnetoresistance is strongly suppressed with increasing pressure when superconductivity becomes predominant. (**d**) The real part of the a.c. susceptibility versus temperature at different pressures.

**The dome-shaped superconductivity behaviour**. We also carried out measurements of resistance around $T_c$ for various external magnetic fields. As seen in Fig. 3a, the zero-resistance state at 24.6 GPa gradually lifted with increasing field, resulting in a decrease in $T_c$. This gave complimentary evidence of the superconducting transition. A magnetic field of 1.5 T almost smears out the superconducting transition. Deviating from the Werthamer–Helfand–Hohenberg theory based on the single-band model, the upper critical field ($H_{c2}(T)$) of WTe$_2$ has a positive curvature close to $T_c$ ($H=0$) as shown in Fig. 3b. This is similar to the case for NbS$_2$ and NbSe$_2$ (refs 22,23). As shown in Fig. 3b, our experimental curve of $H_{c2}(T)$ can be well approximated by a simple relationship of the form $H_{c2}(T) = H_{c2}^{*}(1 - T/T_c)^{1+\alpha}$ (ref. 24), where the estimated value of $H_{c2}(0)$ (2.72 T) is similar to that of NbS$_2$ and NbSe$_2$ (refs 22,23). It is also worth noting that our estimated value of $H_{c2}(0)$ is well less than the Bardeen, Cooper and Schrieffer (BCS) weak-coupling Pauli limit.

The dome-like evolution of $T_c$ was constructed based on the pressure-dependent transport data shown in Fig. 2. It is clear that $T_c$ increased up to a pressure of 16.1 GPa in run no. 1. In the second run, $T_c$ first increased and later decreased with increasing pressure up to a pressure of 68.5 GPa, with a maximum $T_c$ of 7 K at 16.8 GPa. Slight discrepancy can be seen between runs 1 and 2. This is reasonable after considering the situation that the sample size is small (200 × 40 × 5 μm). It is very hard for us to keep the $ab$ plane of WTe$_2$ ideally parallel to the diamond culet. A small deviation from the parallel configuration will cause the pressure applied in different directions. We repeated the experiment using the pressure-transmitting medium of Daphne oil in the third run, which revealed the entire $T_c$ dependence against pressure as shown in Fig. 5a. Here we see the $T_c$–$P$ phase diagram, where $T_c$ starts at 2–3 K at a pressure of 2–4 GPa and increases up to a maximum $T_c$ of 7 K. A dome-shaped superconducting phase is clearly evident. We note the discrepancy between runs 2 and 3 at high pressure. It is well known that measurements using a transmitting medium are normally regarded as quasi-hydrostatic pressure applications, while measurements made without a transmitting medium are regarded as being under uniaxial pressure. The different pressure environments may contribute to some discrepancies[25–27], as observed here.

**The interpretation by theoretical calculations**. We carried out density functional theory (DFT) calculations to better interpret the physics of the observed superconductivity. On the basis of the experimental lattice parameters shown in the Supplementary Table 1 and our optimized internal atomic coordinates, we performed calculations on the electronic structure of WTe$_2$. Due to strong hybridization, the W-5$d$ and Te-5$p$ bands were found to be highly mixed and distributed mainly in the energy range from −6.5 to 3 eV, while all other bands made only a negligible contribution as shown in Supplementary Figs 1 and 2. Crystal field splitting was found to be very small and all the Te-5$p$ and W-5$d$ electrons participate in the electronic states near the Fermi surface. The band structure of WTe$_2$ under ambient conditions is anisotropic with slight dispersion along the Γ-$z$ and greater dispersion along the in-plane directions. Consistent with previous studies[1,3], our calculation also shows that WTe$_2$ is a semimetal with quite small $N(E_F)$ (0.497 states per eV per unit cell), which may explain why WTe$_2$ does not exhibit superconductivity even down to 0.3 K at ambient pressure[2].

We performed total energy calculations for a number of different volumes to simulate the high pressure conditions in WTe$_2$. We optimized the lattice parameters and all the independent internal atomic coordinates for each volume. The obtained volumes versus the total energy behaviour were found to be in good agreement with the Murnaghan equation of state as shown in the Supplementary Fig. 3 (ref. 28). Our theoretical equilibrium unit cell volume (314 Å$^3$) is only about 2.6% larger than the experimental value (306 Å$^3$). Such deviation exists normally in generalized gradient approximation calculations. Our numerical bulk modulus at equilibrium $B_0$ was 56 GPa, slightly larger than that of MoS$_2$ (ref. 29). In Fig. 4a, we show the pressure dependence of the lattice parameter and the $c/a$ ratio. The 2D nature of this compound is clearly exemplified from the different rates of compressibility along the $c$ axis and in the $ab$ plane. The numerical $c/a$ ratio first decreases with pressure until a minimum value is reached at 30 GPa. An upward shift can then be seen, similar to the case for MoS$_2$ (ref. 17). In MoS$_2$, this abnormality was attributed to the occurrence of an isostructural phase transition[17].

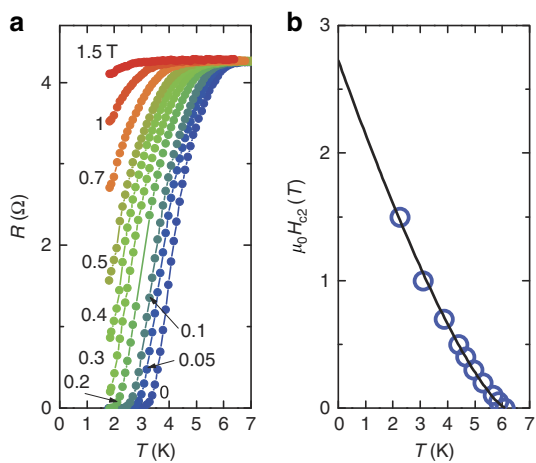

**Figure 3 | The upper critical field analysis of the WTe$_2$ superconductor.**
(**a**) Temperature dependence of the resistance under different fields up to 1.5 T at 24.6 GPa. (**b**) The $T_c$–$H$ phase diagram at 24.6 GPa. The black curve is the best-fit line.

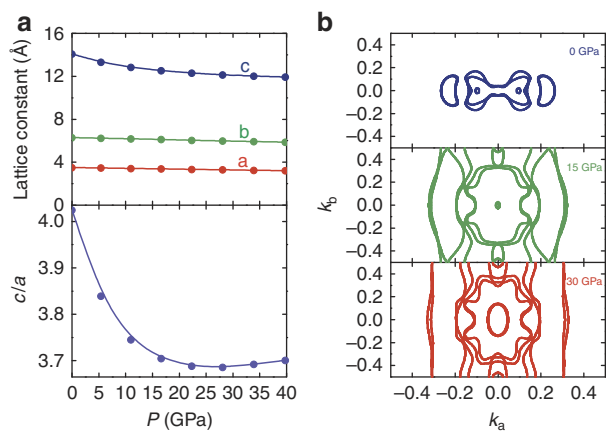

**Figure 4 | Density functional theory calculations.** (**a**) The pressure dependence of the lattice parameters (upper) and $c/a$ ratio obtained from geometry optimization (lower). (**b**) The calculated evolution of the Fermi surface contour at various pressures (marked in the plot).

Our numerical phonon spectrum at zero pressure is in good agreement with the results of Raman spectroscopy measurements[30,31] as shown in the Supplementary Table 3. We also investigated the pressure-induced phonon evolution, suggesting that a possible structural instability may occur under compression, again as shown in the supplementary Fig. 4. The phonon softening found in the finite-displacement method phonon calculation may be related to the $c/a$ abnormality found in the structural optimization as shown in Fig. 4a.

We now turn our attention to the electronic structure under compression. We show the numerical Fermi surface at the $K_z=0$ plane in Fig. 4b. Our zero-pressure Fermi surface is slightly different from the previous one, as shown in the upper plot of Fig. 4b. There is a small Fermi pocket along the $\Gamma$-$y$ direction in our results, which was absent in some previous studies[1–3]. Applying pressure enlarges the hybridization and increases the bandwidth as shown in the Supplementary Figs 1 and 2. Consequently, the size of the electron and hole pockets at ambient pressure increases. Moreover, high pressure also introduces additional Fermi pockets as demonstrated in Fig. 4b. Thus, $N(E_F)$ increases rapidly with pressure as shown in Fig. 5b. As discussed above, applying pressure increases the phonon frequencies, thereby increasing the Debye temperature $T_\theta$. Figure 5a implies that the presence of superconductivity at around 2.5 GPa, together with the sharp increase in $T_c$, can be explained by the increases in $N(E_F)$ and $T_\theta$. $N(E_F)$ continues to increase with pressure owing to the enlargement of existing pockets as well as the appearance of new pockets, and the decrease of $T_c$ above 16.8 GPa may be related to structural abnormalities.

## Discussion

The MR suppression and the appearance of superconductivity can be considered as follows. The large MR is attributed to the perfect compensation between the opposite carriers, where the balance is too delicate to survive the intense pressures used in our experiments. As seen in the simulation, the applied pressure significantly increases the difference between the hole and electronic Fermi pockets as shown in the Supplementary Fig. 1. At the same time, $N(E_F)$ rapidly increases with increasing pressure as shown in Fig. 5b, which is the essential condition of the onset of superconductivity. The applied pressure tunes the carrier balance and the electronic conditions near the Fermi surface and leads to simultaneous MR suppression and the appearance of superconductivity. Interestingly, $H_{c2}(T)$ usually increases with increasing $T_c$ and decreases with increasing Fermi velocity. The numerical Fermi velocity increases with pressure as shown in the Supplementary Table 4 and Supplementary Note. In combination with the dome-like evolution of $T_c$-$P$ (Fig. 5a), we expect that $H_{c2}(T)$ may show complex pressure-dependent behaviour around the optimized pressure, while at high pressure, $H_{c2}(T)$ decreases with increasing pressure.

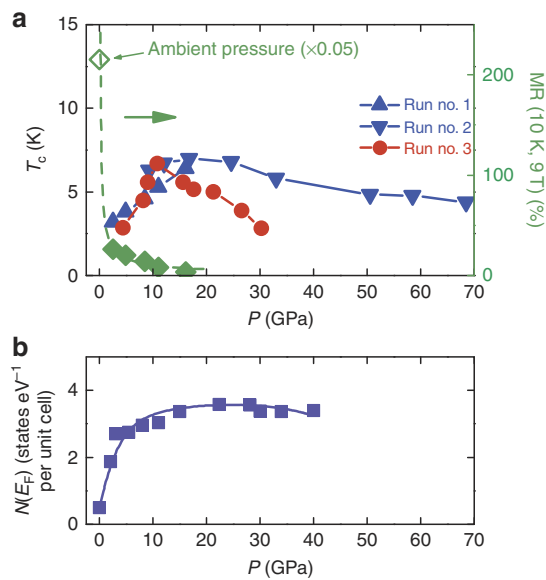

**Figure 5 | The dome-shaped superconducting $T_c$-$P$ phase diagram and possible interpretation.** (**a**) Onset temperature of the superconductivity plotted against applied pressure. A maximum $T_c$ of 7 K occurs near 20 GPa. In runs no. 1 and 2, no pressure-transmitting medium was used while for run no. 3, Daphne oil was used as the pressure medium. The right-hand axis is the magnetoresistance ratio, which is strongly suppressed by the pressure. (**b**) The calculated density of states at the Fermi level plotted against pressure.

In summary, superconductivity was successfully induced in WTe$_2$ by the application of high pressure with a maximum $T_c$ of 7 K at 16.8 GPa. The $T_c$-$P$ phase diagram shows a dome-like superconducting phase, which we attribute to an enrichment of $N(E_F)$ for the low-pressure regime and a possible structural instability at high pressures as suggested by DFT calculations. The semimetal-like electronic dispersion, unsaturated large MR and superconductivity were all observed in WTe$_2$, all contributing to the extremely interesting physics seen in this TMD material.

## Methods

**Crystal growth and characterization.** WTe$_2$ single crystals were grown using a chemical vapour transport technique. Stoichiometric W and Te powders were ground together and loaded into a quartz tube with a small amount of TeBr$_4$ (transport agent). All weighing and mixing was carried out in a glove box. The tube was sealed under vacuum and placed in a two-zone furnace. The hot zone was maintained at 800 °C for 10 days and the cold zone was maintained at 700 °C. A Bruker SMART diffractometer equipped with a charge-coupled device-type area detector was used to determine the crystal structure. The data were collected at room temperature with graphite monochromated Mo-K$\alpha$ radiation ($\lambda=0.71073$ Å). SADABS[32] supplied by Bruker was used to perform the absorption correction. The Patterson method was used to resolve the structure[33], which we then refined using full-matrix least squares on all F$^2$ data with the SHELXL-97 program[32]. All atoms were refined anisotropically.

**Transport measurements and high-pressure experiments.** The resistance data were collected in a screw-pressure-type diamond anvil cell (DAC) made of non-magnetic Cu–Be alloy. The diamond culet was 300 μm in diameter. A T301 stainless-steel gasket was pre-indented from a thickness of 200 to 30 μm, leaving a pit inside the gasket. A hole with a diameter of 280 μm was drilled at the centre of the pit using laser ablation. The pit of the indented gasket was then covered with a mixture of epoxy and fine cubic boron nitride (cBN) powder and compressed firmly to insulate the electrode leads from the metallic gasket. For runs 1 and 2, we used the standard four-probe method to obtain the resistance measurements. The cBN-covered pit served as a sample chamber, into which a $WTe_2$ single crystal of dimensions $200 \times 40 \times 5$ μm was inserted without a pressure-transmitting medium. For run 3, a hole with diameter of 100 μm was further drilled at the centre of the cBN-covered pit, and then a single-piece sample with dimensions $40 \times 40 \times 5$ μm was loaded simultaneously using Daphne 7373 oil as the pressure-transmitting medium. We used van der Pauw-like topology to arrange the four probes. The current was introduced into one side and the drop in voltage along the other side was recorded. Some ruby powder at the top of the sample for runs 1 and 2 and a ruby ball next to the sample for run 3 served as pressure markers. The pressure was determined using the ruby fluorescence method at room temperature. Platinum (Pt) foil with a thickness of 5 μm was used for the electrodes. The gasket surface outside the pit was insulated from the electrode leads using a layer of Scotch tape. The DAC was placed inside a homemade multifunctional measurement system (1.8–300 K, JANIS Research Company Inc.; 0–9 T, Cryomagnetics Inc.) with helium (He) as the medium for heat convection to obtain a high efficiency of heat transfer and precise temperature control. Two Cernox resistors (CX-1050-CU-HT-1.4L) located near the DAC were used to ensure the accuracy of the temperature in the presence of a magnetic field. The ambient pressure electrical transport were carried out in a Cryomagnetics cryostat with an SR830 (Stanford Research Systems) digital lock-in amplifier. Ohmic contacts were made using gold wires and silver paste.

The a.c. susceptibility was measured using a magnetic inductance technique based on the ideal diamagnetism of a superconductor. One signal coil was wound around a diamond tip with an additional identical compensating coil tightly adjacent to it. These two coils are known as a pick-up coil. Outside the pick-up coil was an exciting coil, into which the a.c. current was fed with a magnitude of 100 μA and a frequency of 997 Hz. The diamond culet was 800 μm in size and a Be–Cu gasket was pre-indented from 450 to 150 μm. A piece of $WTe_2$ single crystal sample with dimensions $500 \times 500 \times 20$ μm was inserted into the pre-indented Be–Cu gasket chamber without a pressure-transmitting medium. Ruby powder placed at the top of the sample served as a pressure marker. The pressure was determined using the ruby fluorescence method at room temperature.

**Density functional calculations.** The electronic structure was calculated based on DFT as implemented in the Vienna *ab initio* simulation package code[34,35]. The Perdew–Becke–Erzenhof parameterization of the generalized gradient approximation was adopted as the exchange-correlation function[36]. A plane-wave basis set was employed within the framework of the projector augmented wave method[37] and the cutoff energy of 500 eV had been tested to ensure it was sufficient for convergence. The Brillouin zone was sampled using the Monkhorst-Pack method[38] with a **k**-point grid $14 \times 7 \times 3$. The relaxations of cell geometry and atomic positions were carried out using a conjugate gradient algorithm until the Hellman–Feynman force on each of the unconstrained atoms was $<0.001$ eV Å$^{-1}$. The convergence criterion of the self-consistent calculations was $10^{-5}$ eV between two consecutive steps. We adopted the Gaussian smearing scheme[39] using a smearing width of 0.05 eV. We also included spin-orbital coupling in our calculations.

The calculations of phonon spectrum were performed in a $3 \times 2 \times 1$ supercell, with interatomic forces being computed using the Vienna *ab initio* simulation package code code with the small displacements method[40]. From these, force-constant matrices and phonon frequencies were extracted using the PHONOPY code[41]. The cutoff energy of 500 eV and the Gaussian smearing method with a 0.05 V smearing width were used in the phonon calculations. Our numerical calculations showed that the phonon frequency difference between k-meshes of $3 \times 3 \times 2$ and $5 \times 4 \times 4$ was quite small. Hence, a $3 \times 3 \times 2$ **k**-point grid was used for the phonon calculations for all the pressures. The effect of the exchange-correlation function pseudopotential, cutoff value and smearing value were also carefully checked. It is worth mentioning that our band structure was consistent with previous results and our numerical Raman frequencies were also in good agreement with the experimental results. These again justify the quality of our numerical electronic structure and phonon results.

### Acknowledgements
We gratefully acknowledge the financial support of the National Key Projects for Basic Research of China (Grant Nos: 2013CB922103, 2011CBA00111 and 2011CB922103), the National Natural Science Foundation of China (Grant Nos: 91421109, 11134005, 61176088, 11222438, U1332143, 51372249, 11374307, 11374137, 91122035, 11174124 and 11274003), the PAPD project, the Natural Science Foundation of Jiangsu Province (Grant BK20130054), and the Fundamental Research Funds for the Central Universities. We would also like to acknowledge the helpful assistance of the Nanofabrication and Characterization Center at the Physics College of Nanjing University and Professor D.L. Feng at Fudan University.

### Author contributions
X.W. proposed the work. F.S., Z.Y. and B.W. designed the research. X.P. prepared the sample and made the material characterizations. X.C. carried out the high-pressure resistance and a.c. susceptibility measurements with the assistance of Y.Z. (Yonghui Zhou) and Z.C. and F.Y.. X.W., H.L. and Y.F. carried out the theoretic calculations. X.W., F.S., Z.Y. and X.-C.P. co-wrote the paper. All authors commented on the manuscript.

### Additional information
**Supplementary Information** accompanies this paper at http://www.nature.com/naturecommunications

**Competing financial interests**: The authors declare no competing financial interests.

**Reprints and permission** information is available online at http://npg.nature.com/reprintsandpermissions/

**How to cite this article**: Pan, X-C. *et al.* Pressure-driven dome-shaped superconductivity and electronic structural evolution in tungsten ditelluride. *Nat. Commun.* 6:7805 doi: 10.1038/ncomms8805 (2015).